\documentclass{ijmpi_authors}
\usepackage{color}
\usepackage{graphicx}
\usepackage{amsmath}
\usepackage{subcaption}

\begin{document}

\twocolumn[
\title{Towards a mechanical MPI scanner based on atomic magnetometry}

\author{Colombo}{Simone}{a,\ast}
\author{Lebedev}{Victor}{a}
\author{Tonyushkin}{Alexey}{b}
\author{Gruji\'c}{Zoran~D.}{a}
\author{Dolgovskiy}{Vladimir}{a}
\author{Weis}{Antoine}{a}

\affiliation{a}{Physics Department, University of Fribourg, Chemin du Mus\'{e}e 3, 1700 Fribourg, Switzerland}
\affiliation{b}{Physics Department, University of Massachusetts Boston, Boston, MA 02125 USA}
\affiliation{\ast}{Corresponding author, email: simone.colombo@unifr.ch}

\maketitle

\begin{abstract}
We report on our progress in the development of an atomic magnetometer (AM) based low-frequency Magnetic Particle Imaging (MPI) scanner, expected to be free from Specific Absorption Rate (SAR) and Peripheral Nerve Stimulation (PNS) constraints. 
We address major challenges in coil and sensor design due to specific AM properties. 
Compared to our previous work we have changed the AM's mode of operation towards its implementation for detecting the weak magnetic field produced by magnetic nanoparticles (MNP) in the presence of nearby-located strong drive/selection fields.
%
%
We demonstrate that a  pump-probe AM scheme in a buffer gas filled alkali vapour cell can tolerate mT/m gradients while maintaining a sensitivity in the one-digit pT/$\mathrm{\sqrt{Hz}}$ range over a bandwidth from DC to several ~kHz. 
We give a detailed description of the drive/selection coils' geometry and their hardware implementations that provides field-free-line (FFL)  operation, compatible with a best performance AM operation. 
We estimate the achievable field of view and spatial resolution of the scanner as well as its sensitivity, assuming mechanical scanning of a Resovist sample through the field-free point/line.
\end{abstract}
]
\section{Introduction}\label{ref:intro}
Since its invention in 2005 \cite{GleichN2005} Magnetic Particle Imaging (MPI) has developed into a  mature technology with a number of variants and designs\cite{Knopp2012,GoodwillAM2012,WeizeneckerAPD2008}. 
All MPI implementations are based on the detection of the MNP response to an oscillating drive field produced by transmit coils.
The detection by resonant (or non-resonant \cite{Tay2016}) receive coils relies on Faraday's induction law, implying a voltage signal proportional to the drive frequency.
The detection efficiency of the oscillating MNP magnetization $M(t)$ used to encode the sample's spatial density distribution is thus favoured for high frequency excitation.
However, concerns have been raised about the specific absorption rates, which depend on the drive field's amplitude and frequency in a similar manner as the MNP response proper \cite{Saritas2013}.
This feature limits the frequency and  amplitude of the drive field, thus affecting both the sensitivity and spatial resolution of the method.

Replacing  the pick-up coil with a sensitive magnetic field sensor with a  flat frequency response (down to DC) may circumvent the above mentioned limitations.
An analysis of the performance of  state-of-the-art MPI systems shows that this alternative detection method calls for a magnetometer with magnetic sensitivity in the lower  pT range, preferably reaching 100 fT, and having a flat frequency response up to several 10~kHz.
Following our successful demonstration that magnetic particle spectra (MPS) can be recorded with a state-of-the-art atomic magnetometer (AM) \cite{colombo2016a,colombo2016b}, we are in the process of developing an AM-based MPI system aiming at a competitive sensitivity and resolution, while deploying drive frequencies in the lower (or sub-) kHz frequency range. 
Atomic magnetometers, also known as optical magnetometers or optically pumped atomic magnetometers, measure the magnetic field by the optical readout of  the Larmor precession frequency  
\begin{equation}
f_\mathrm{L}=\gamma |\vec{B}|=\gamma B\quad
\left(\gamma=\frac{\mu_B}{h(I+1/2)}{\approx}3.5~\mathrm{Hz/nT}\right)\,,
\label{eq:omegaL}
\end{equation}
of spin-polarized paramagnetic atoms in the magnetic flux density $B(t){\propto}M(t)$ of interest.
In Eq.~\eqref{eq:omegaL} $\mu_B$ is the Bohr magneton, $h$ Planck's constant, and $I{=}7/2$ the nuclear spin of the $^{133}$Cs atoms used in the magnetometer.
We note that $B$ represents the average flux density in the intersection volume $V$ of the pump and probe beams (see Fig.~\ref{fig:opm_1}).
An AM sensor is a sealed glass bulb containing a vapour of the sensing atoms,  $^{133}$Cs in our case.
Circularly-polarized resonant laser light (pump beam) produces the required spin polarization.
A second beam (probe beam) -- derived from the same or another laser -- reads out the spin precession that is impressed by a magnetic resonance process as a power or polarization modulation onto the transmitted probe beam.

The sensitivity of the AM scales with the amount of sensing atoms, i.e., with the volume $V$, but degrades when the flux density changes over the sensed  volume, since inhomogeneous $B$-fields broaden the magnetic resonance line.
The AM deployed in our initial MPS experiments \cite{colombo2016a,colombo2016b}  could reach a sensitivity of $200~\mathrm{fT/\sqrt{Hz}}$ in a 2~kHz bandwidth under optimal conditions, i.e. in a homogeneous field inside state of art magnetic shielding.
However, such AM originally designed for the fT-performance under shielded conditions, stopped working in the ambient field gradients above $0.2~$mT/m, a value much smaller than typical fringe field amplitudes in the vicinity of any MPI/MPS coils.
%
Demand for high sensitivity in presence of strong field gradients lead us to the development of self-compensating solenoids for MNP magnetization, reducing the fringe drive field at the sensor position by a factor of 10$^4$ compared to the field at the sample position (in presence of a $1$~T/m gradient at the sample position).
In that geometry the magnetometric sensitivity was reduced to a few $\mathrm{pT/\sqrt{Hz}}$, and the drive coil design did not allow us to convert the system into a simple MPI scanner.

Here we describe our recent improved design of the experimental set-up, in particular the sensor and coil, in view of developing an operational 2D MPI scanner and its future upgrade to a volumetric scanner.
%

%
\section{Optical magnetometer}\label{ref:AM}
\begin{figure}[!ht]
	\centering
	\includegraphics[width=\columnwidth]{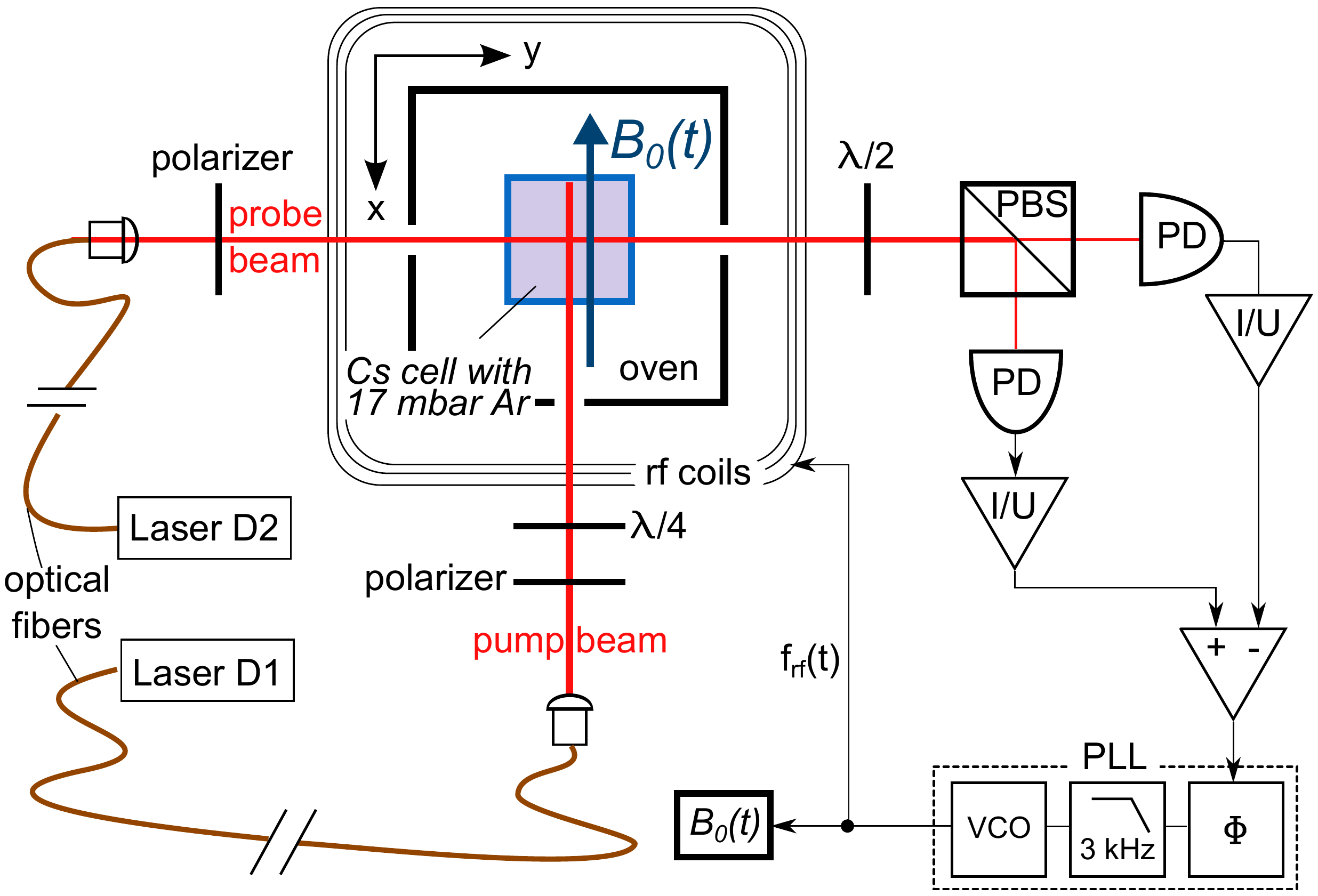}
	\caption{Sketch of the magnetometry part of the apparatus with polarizing beam splitter (PBS), photodiode (PD), current--to--voltage converter (I/U), and voltage-controlled oscillator (VCO).}
	\label{fig:opm_1}
\end{figure}

The operation of our magnetometer is sketched in Fig.~\ref{fig:opm_1}.
%
%
The circularly-polarized pump laser beam ($\lambda\sim$894~nm) is resonant with the $4{\rightarrow}3$ hyperfine component of the $6S_{1/2}{\rightarrow}6P_{1/2}$ ($D_1$) atomic transition spin-polarizes the cesium atoms by optical pumping, see Chapter 4 in~\cite{budke13}.
The probe laser beam is linearly-polarized and its polarization is analyzed by a balanced polarimeter.
The system is operated in a magnetically unshielded environment, in which the local laboratory field is compensated and an offset field $B_0$ of $\approx$27~$\mu$T (corresponding to a Larmor frequency of $\approx$100~kHz) along $x$ is applied to the atoms.
%
%

A weak radio frequency magnetic field (rf-field) oscillating along the $z$ direction is used to resonantly drive the atomic spin polarization produced by the pump beam on a cone around the $x$ axis, leading to an oscillating component polarization along the probe beam direction.    
This oscillating component induces a corresponding oscillation of the direction of the probe beam's linear polarization that is detected by the balanced polarimeter.
A phase  detector (marked $\Phi$) and a voltage controlled oscillator (VCO) drive the rf-coil.
When exposed to a time-independent field $\vec{B}_0{\parallel}\hat{x}$, the oscillation frequency of this phase-locked loop is proportional to $B_0$, following Eq.~\eqref{eq:omegaL}.
Any component $\delta B_x(t)$ of a time-dependent field along $\hat{B}_0$ -- such as the one produced by the harmonically driven MNP sample -- will thus induce a frequency modulation, whose amplitude ($\propto \delta B_x$) can be extracted by a suitable demodulation technique \cite{colombo2016b}. 

We stress that only field components $\delta\vec{B}$  parallel to $\vec{B}_0$ yield a linear response, since  they change the Larmor frequency in a linear manner, while transverse components yield only second order corrections.  
\begin{figure}[!hb]
	\centering
	\includegraphics[width=.99\columnwidth]{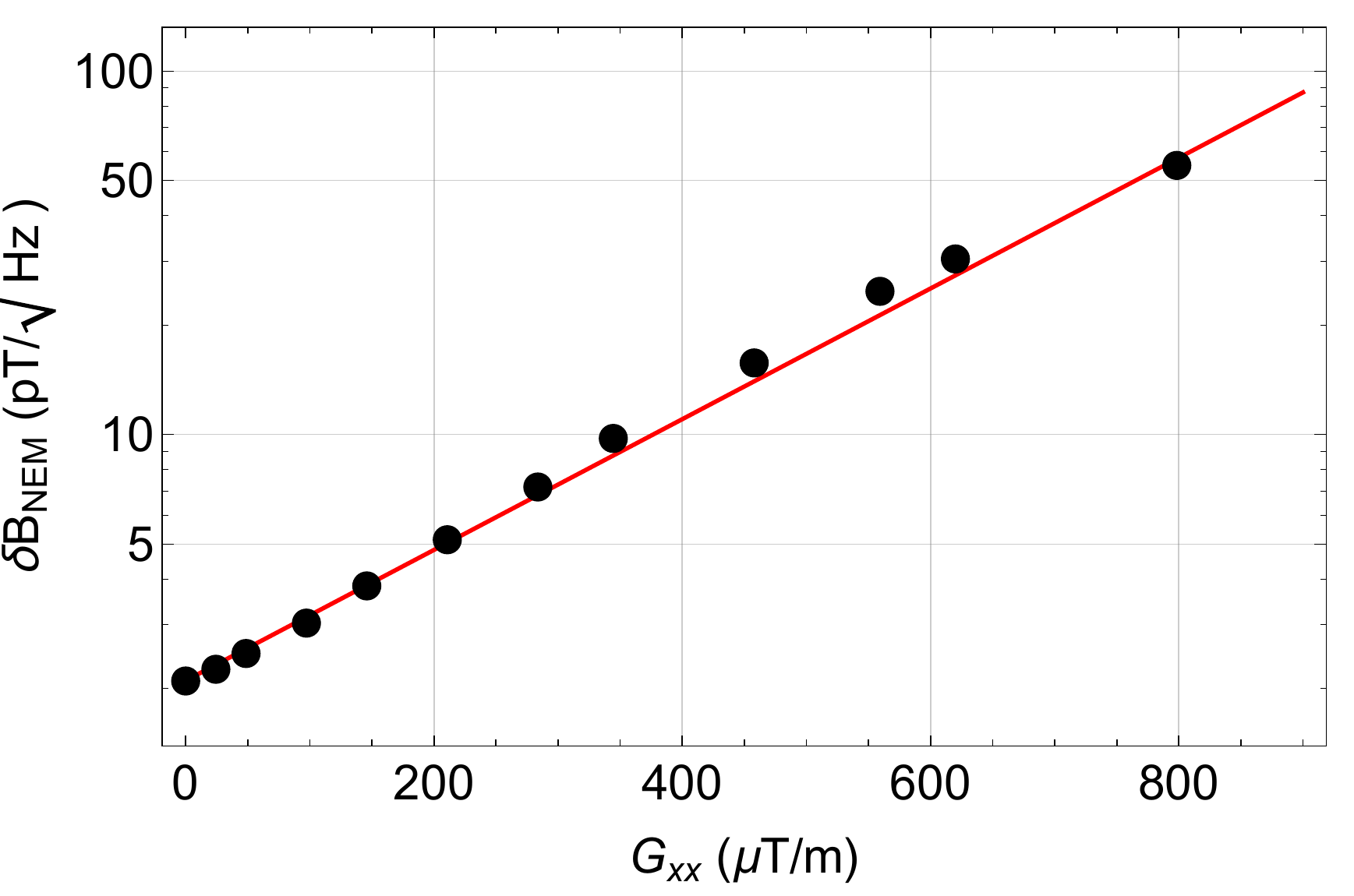}
	\caption{Magnetometer sensitivity (defined as noise-equivalent magnetic flux density, NEM) as a function of field gradient in the sensor volume. }
	\label{fig:3opm_gradient}
\end{figure}

To improve the performance of our AM in the presence of field gradients, the sensing volume must have the smallest possible size, such that magnetic resonance line broadening due to field variations over the volume is minimized.
Argon buffer gas at 17~mbar limits the diffusive motion of the spin-polarized Cs atoms to the region illuminated by the pump laser beam.
The sensing volume is then further constrained by having the (orthogonally propagating) pump and probe beams intersect in a volume of ${\approx}2{\times}2{\times}2~\mathrm{mm}^3$.
Since the magnetometric sensitivity depends on the number of contributing atoms, the small sensing volume implies the need to increase the atomic density, itself proportional to the saturated vapor pressure.
For this reason the cell is installed inside a miniature oven ( ${\approx}7{\times}7{\times}5~\mathrm{cm}^3$) heated to an optimized temperature of $55^\circ$~C.

We have tested the performance of the AM in presence of field gradients.
For this we exposed the AM to a quadrupole field with a linear gradient
$G_{xx}{=}\mathrm{d}B_x/\mathrm{d}x$ along the offset field $\vec{B}_0$.
The results are shown in Fig.~\ref{fig:3opm_gradient}.
%

%

\section{Coil design}
Designing the  coils  for operating  an  MPI scanner based on atomic magnetometry is a very delicate task.
The selection coils should produce a gradient on the order of T/m at the sample position, while the fringe field of that coil at the AM position must be as small and homogeneous as possible in order  to achieve an optimal sensitivity.
Figure~\ref{fig:3opm_gradient} shows that in a gradient $G_{xx}$ of 500~$\mu\mathrm{T/m}$ our magnetometer has a sensitivity $\delta{B_\mathrm{NEM}}$ of $\approx 20\mathrm{~pT/\sqrt{Hz}}$.
When aiming at a gradient of 1~T/m at the MNP sample position, one has thus to insure that the stray gradient `seen' by the magnetometer is suppressed by a factor of at least 2'000 with respect to that value.
%

On the other hand one needs to ensure that the detected modulated field component $\delta B_x(t)$ produced by the MNP sample in the sensor is not significantly  perturbed by a fringe field component of the modulation field $H^\mathrm{mod}(t)$.

\begin{figure}[!ht]
	\centering
	\includegraphics[width=\columnwidth]{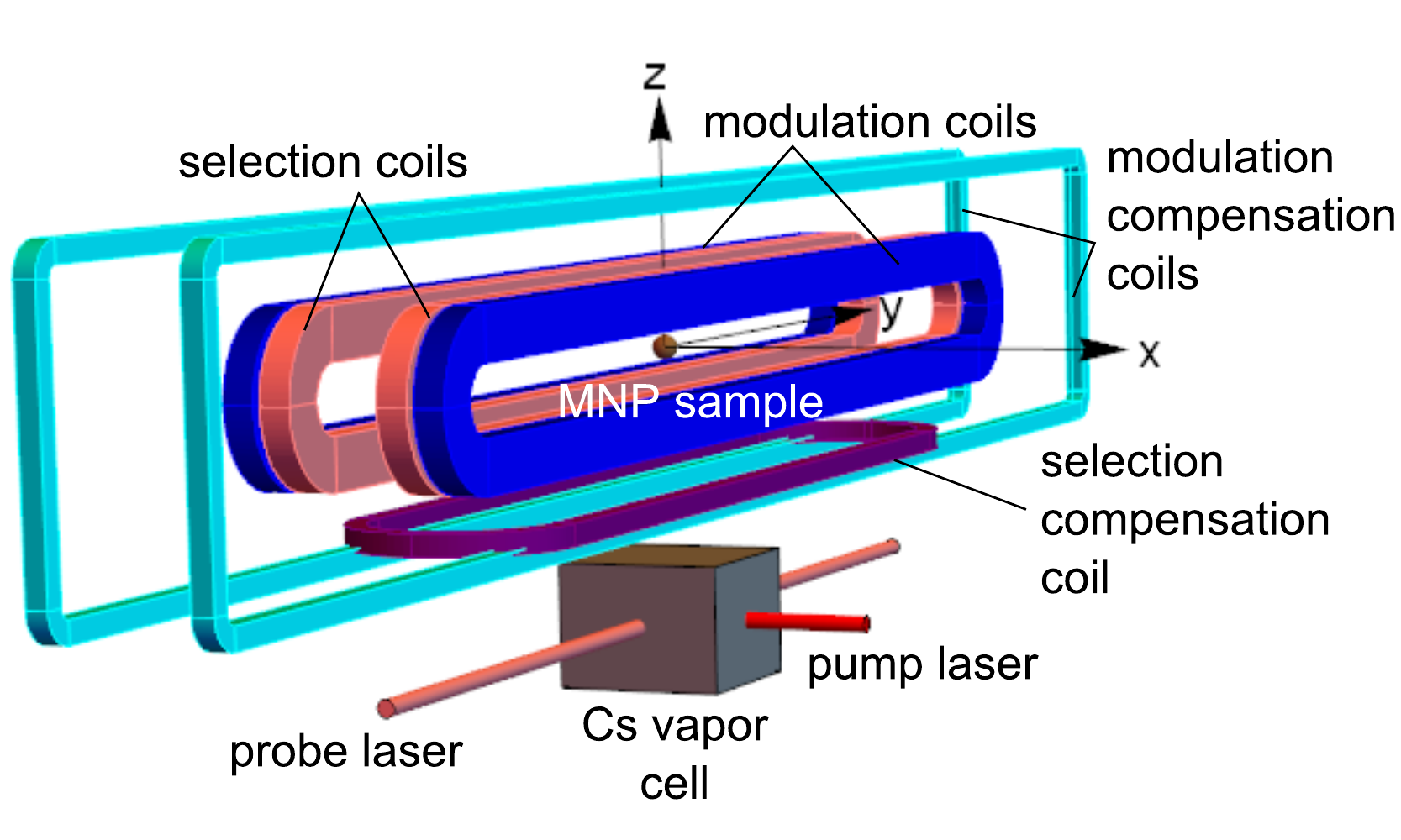}
	\caption{Sketch (to scale)  of the coil design for the  proposed MPI scanner. 
Race-tracks represent different current-carrying coils: pink -- the selection coils, purple -- the selection compensation coil, blue -- the modulation coils, and cyan -- the modulation compensation coils. 
    The directions of current flow  are given in Fig.~\ref{fig:2dsketch}.}
	\label{fig:coils3d1}
\end{figure}

We produce the MNP selection and modulation fields by means of elongated coils \cite{Tonyushlkin2010,Tonyushlkin2016} shown in Fig.~\ref{fig:coils3d1} (selection coils in light-red and  modulation coils in blue).
Each of the 300~mm long coils consists of 39 layers of copper tape with a 6.25$\times$0.25~mm$^2$ cross-section, isolated by a 25~$\mu$m thick Kapton insulator on one side.
The vertical extension of the coil is 40~mm yielding an aspect-ratio larger than 7.
The coils have the advantage of  being mechanically very stable and sustaining a large current density  without significant heating.

\begin{figure}[!b]
\includegraphics[width=.99\columnwidth]{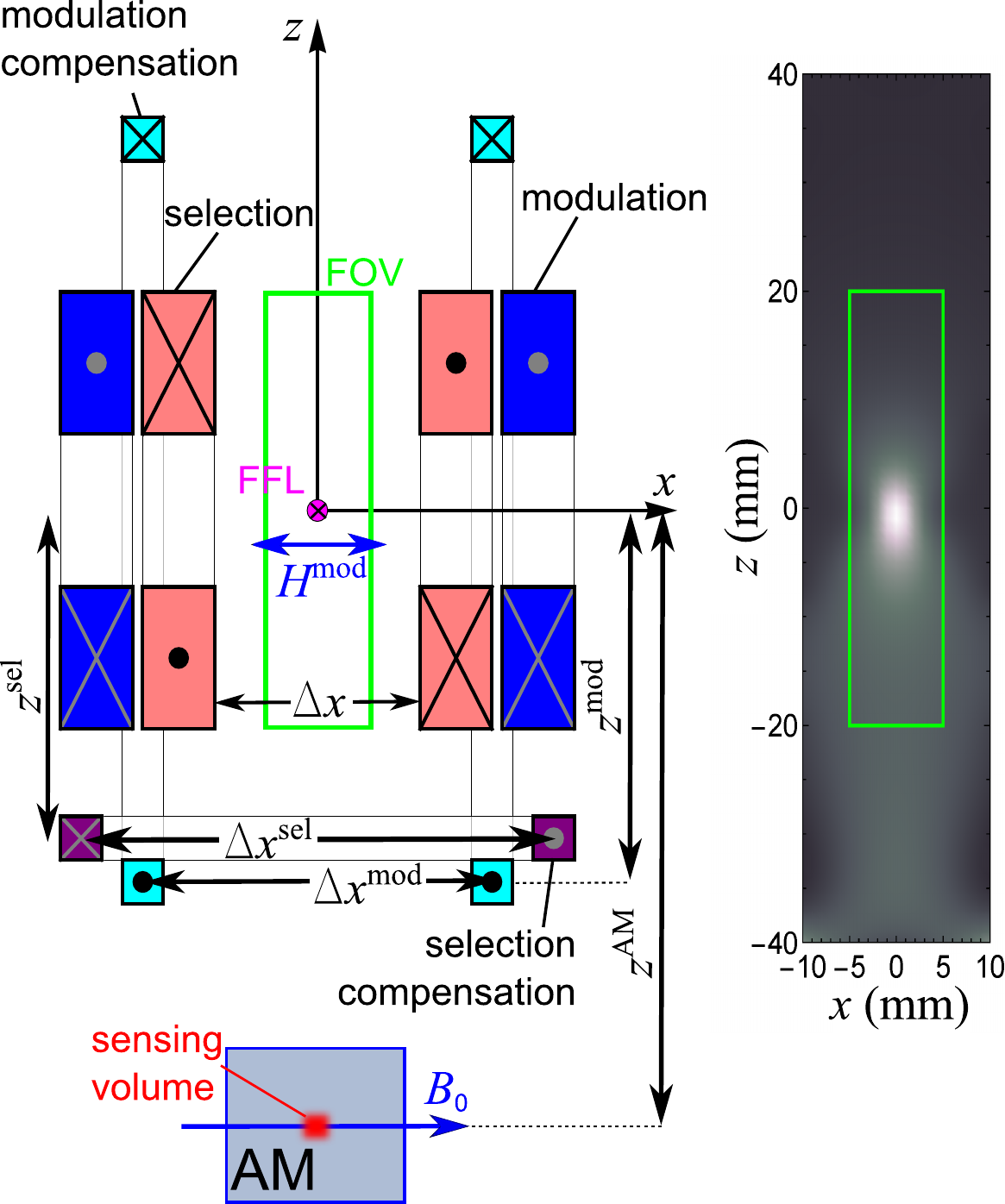}
\caption{Left: Cross-sectional view (to scale) of the deployed coils.
The green rectangle indicates the field of view (FOV) in the $y{=}0$ plane limited by (not shown) mechanical components.
Right: Anticipated point spread function (image produced by a point-like sample) in a gradient $\mu_0 G_{xx}=H_k/\mathrm{mm}$.
The green rectangle delimits the FOV.
}
\label{fig:2dsketch}
\end{figure}

The selection coils are placed in an oppositely-poled configuration (quadrupole field) which creates a field free line (FFL) extending along the y-axis.
For an aperture $\Delta x=20$~mm (defined in Fig.~\ref{fig:2dsketch} and allowing a geometrically accessible 10$\times$40~mm$^2$ field of view, FOV) the gradient scales with current as 
\begin{equation}
\frac{\mathrm{d}G^\mathrm{sel}_{xx}}{\mathrm{d}I}=-\frac{\mathrm{d}G^\mathrm{sel}_{zz}}{\mathrm{d}I}=0.1~\frac{\mathrm{T}}{\mathrm{m}}\, \mathrm{A}^{-1}\,.
\end{equation}
%

\begin{figure}[!t]
	\centering
	\includegraphics[width=\columnwidth]{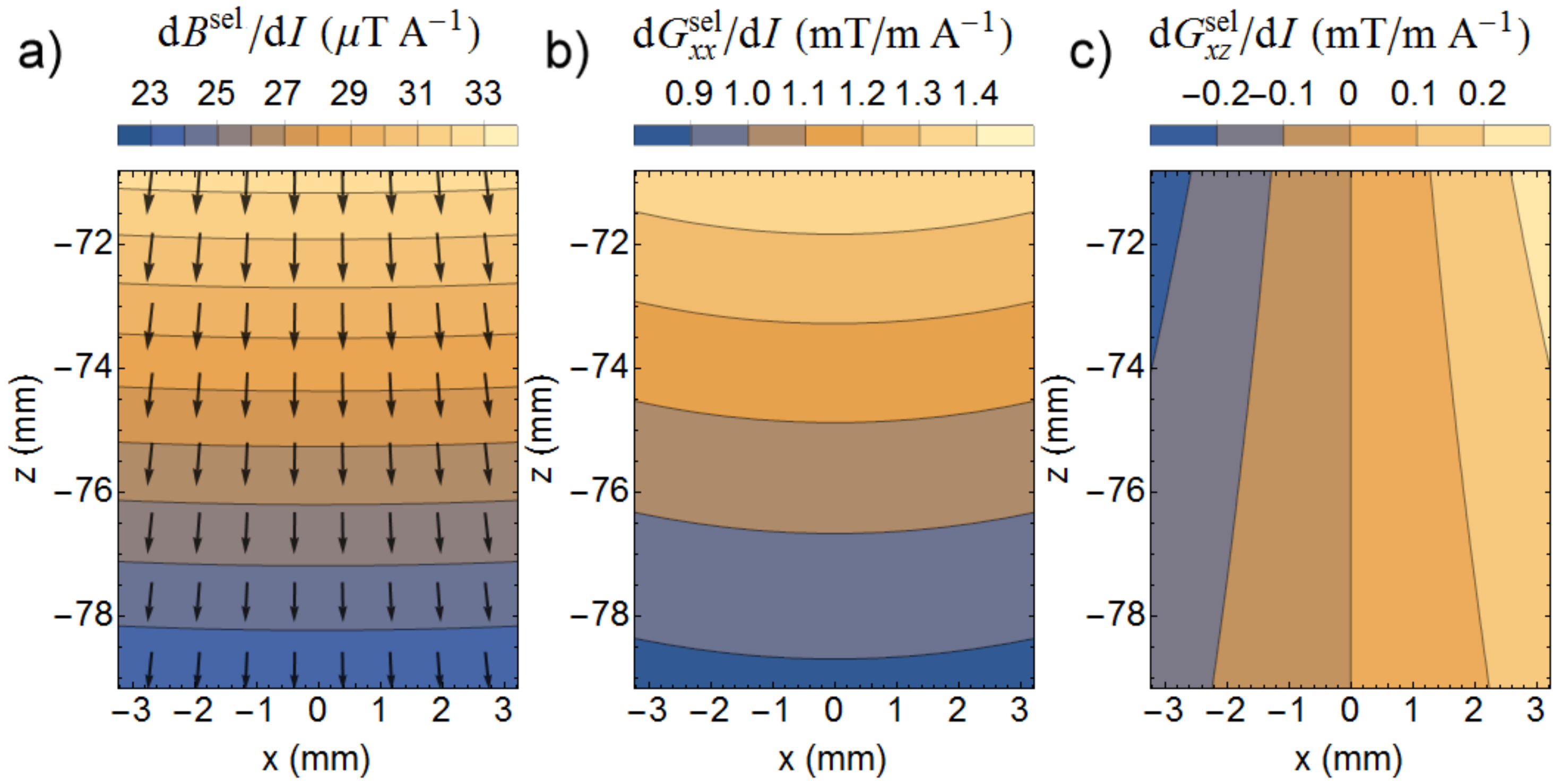}
	\caption{Simulated fringe field from the \emph{selection coils} (a) and its two gradient components $G_{xx}$ (b) and $G_{xz}$ (c) at the AM position, located in the $y$=0 plane, where  $B_y=0$.}
	\label{fig:leak_Sel}
\end{figure}

At the magnetometer, located at $z^\mathrm{AM}{=}-$75~mm below the FFL, the selection coils produce the fringe field pattern shown in Fig.\ref{fig:leak_Sel}.a.
The graphs $b$ and $c$ of that figure show the relevant gradients that influence the AM sensitivity.
The field patterns produced by the selection coils at both the sample and the AM positions are  well approximated by the field from four infinitely long rods carrying the same total current and having the same cross-section as the actual coils.
The 10~A current needed to produce a gradient of $G_{xx}^\mathrm{sel}(z^\mathrm{sample})$ of $\approx$1~T/m at the MNP sample position produces a $B^\mathrm{sel}_z$ field component of $\approx$260~$\mu$T and a gradient $G_{xx}^\mathrm{sel}(z^\mathrm{AM})$ of ${\approx}11$~mT/m at the AM position.

For the above reasons we need compensation coils that suppress  the fringe fields from both the selection coil and the modulation coil at the AM position.

\subsection{Selection field compensation}

In a simulation calculation we have tuned the aspect ratio and the position of the coil (shown in purple in Fig:\ref{fig:coils3d1}) that compensate the selection field's fringe field at the magnetometer position.
The tuning criteria are the simultaneous minimization of the gradients $G^\mathrm{sel}_{xx}$, $G^\mathrm{sel}_{xz}$  as well as the $B^\mathrm{sel}_z$ field component at the sensor position.
In order to perform this tuning we have fitted the field component $B_z$ and the gradients $G_{xx}$ and $G_{xz}$ induced by two infinitely long wires with oppositely flowing current to the corresponding fringe field/gradients of the selection coils.
The parameters of this fitting procedure are the spacing $\Delta{x}^\mathrm{sel}$ between the wires, the vertical position $z^\mathrm{sel}$ and the ratio $\alpha^\mathrm{sel}$ between the \emph{total} selection compensation current and the selection coil current.
Reversing the compensation current then yields a field pattern that locally compensates $B^\mathrm{sel}_z$, $G^\mathrm{sel}_{xx}$ and $G^\mathrm{sel}_{xz}$ leaking from the selection coils.  
The infinitely long rods used in the modeling transfer to the real world as multiple loops of copper wire wound on a racetrack support having $\Delta{x}^\mathrm{sel}$ as aperture in the $x$-direction and an extension $\Delta{y}^\mathrm{sel}$ in the $y$-direction.
For an aperture $\Delta{x}=20$~mm of the selection coils we obtain the following parameters for the selection compensation coil $\Delta{x}^\mathrm{sel}=42.5$~mm, $\Delta{y}^\mathrm{sel}=280$~mm, $z^\mathrm{sel}=-39$~mm and $\alpha^\mathrm{sel}=5.6$.

\subsection{Modulation field compensation}

\begin{figure}[!t]
	\centering
	\includegraphics[width=\columnwidth]{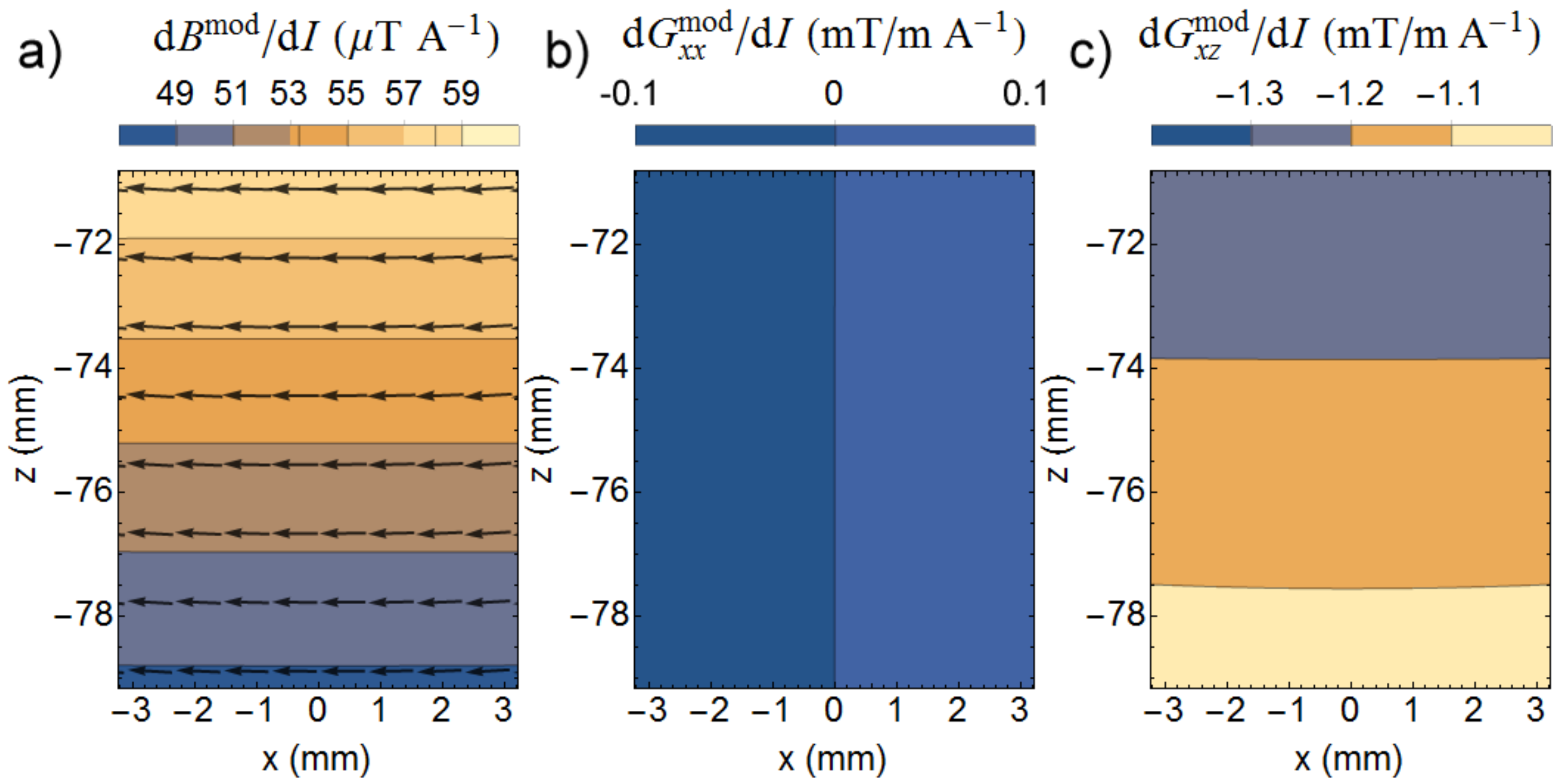}
	\caption{Simulated fringe field amplitude from the \emph{modulation coils} (a) and its two gradient components $G_{xx}$ (b) and $G_{xz}$ (c) at the AM position, located in the $y$=0 plane, where  $B_y=0$.}
	\label{fig:leak_Mod}
\end{figure}

The modulation coils (shown in blue in Figs.~\ref{fig:coils3d1},\ref{fig:2dsketch}) produce an oscillating homogeneous field oriented along the $x$-axis at the MNP sample position.
For the geometry shown in Fig.~\ref{fig:2dsketch} and an aperture $\Delta{x}$ of $20$~mm, the field's amplitude produced by the modulation coils at the sample position scales with current as d$B^\mathrm{mod}_x$/d$I$=0.76~mT\,A$^{-1}$.
At the sensor position the modulation coils produce the field amplitude pattern shown in Fig.\ref{fig:leak_Mod}.a.
The gradients $G^\mathrm{mod}_{xx}$ and $G^\mathrm{mod}_{xz}$ leaking from those coils are very small as shown in Fig.\ref{fig:leak_Mod}.b-c .
As with the selection coils, the pattern produced by the modulation coils in the region of interest is well approximated by one of four infinitely long rods with the same cross-section and carrying the same total current as the coils.

The spacing $\Delta{x}^\mathrm{mod}$ between the two modulation compensation coils and their vertical extension are chosen to minimize simultaneously the gradient components $G^\mathrm{mod}_{xx}$ and $G^\mathrm{mod}_{xz}$ as well as the field component $B^\mathrm{mod}_x$.
The procedure follows the one outlined for the selection coils.
%
We fit the component $B_x$ and the gradients $G_{xx}$ and $G_{xz}$ produced -- at the sensor position -- by two infinitely long wires to optimize the spacing $\Delta{x}^\mathrm{mod}$ between the wires, the vertical position $z^\mathrm{mod}$ and the ratio $\alpha^\mathrm{mod}$ between the \emph{total} modulation compensation current and the modulation coil current.
%
%
Reversing the compensation current then minimizes the total field/gradients at the AM position.
In the apparatus the modelled modulation compensation system is realized as two extended rectangular coils (cyan in Fig.\ref{fig:coils3d_1}) with multiple loops.
The vertical extension and spacing of the coils are $2\left|z^\mathrm{mod}\right|$ and $\Delta{x}^\mathrm{mod}$, respectively.
For an aperture $\Delta{x}$ of $20$~mm we get the optimized parameters $\Delta{x}^\mathrm{mod}=27$~mm, $z^\mathrm{mod}=-45.6$~mm and $\alpha^\mathrm{mod}=4.8$.

\subsection{Performance}

The coil system described above allows controlling the selection and modulation fields leading to a low field/gradient region at a distance of $\approx$75~mm  from the sample at levels that do not significantly degrade the magnetometer's sensitivity.

%
%
%
\begin{figure}[!ht]
	\centering
	\includegraphics[width=\columnwidth]{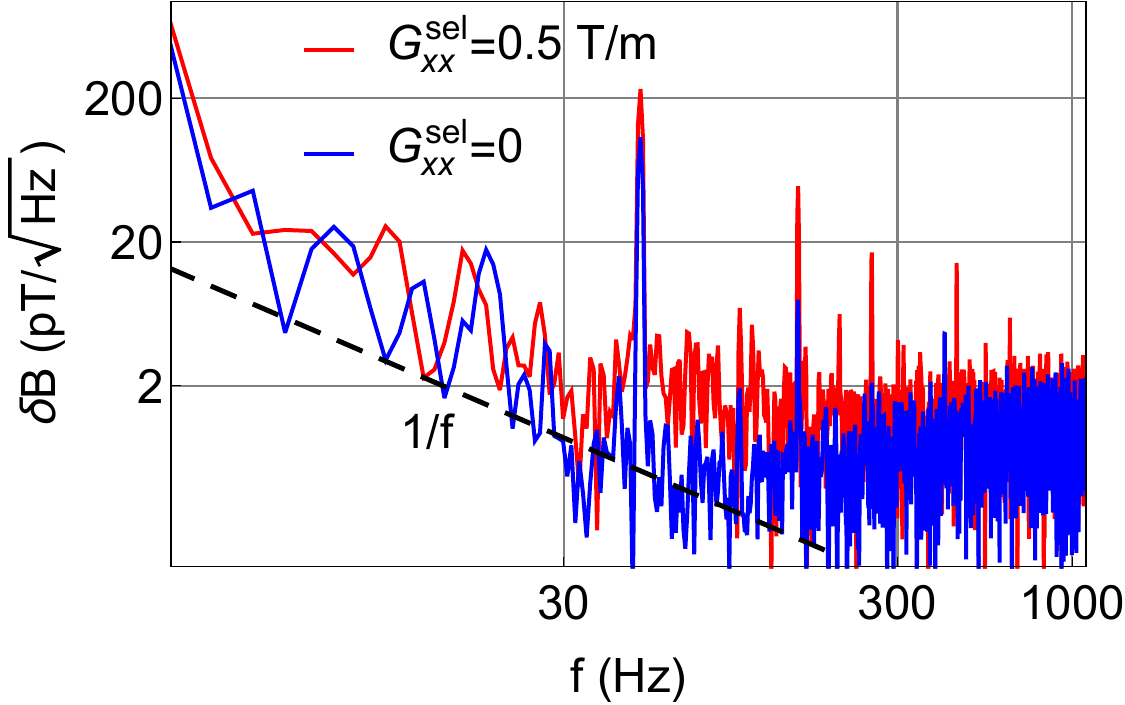}
	\caption{AM sensitivity with and without powered MPI selection coils.
$G^\mathrm{sel}_{xx}$ denotes the gradient at the MNP location for respective plot traces. The dashed line marked `1/f' is meant to guide the eye.}
	\label{fig:dBspectrum}
\end{figure}

We have measured the magnetic field's noise spectral density $\delta{B}_\mathrm{NEM}$ of the AM described in Sec.~\ref{ref:AM} with and without gradient ($G^\mathrm{sel}_{xx}{=}{-}G^\mathrm{sel}_{zz}{=}0.5$T/m) applied at the MNP location.
The respective spectra are shown in Fig.\ref{fig:dBspectrum}.
While the unperturbed magnetometer reaches sub-pT/$\mathrm{\sqrt{Hz}}$ sensitivity in the range of 30-300~Hz and below 2~pT/$\mathrm{\sqrt{Hz}}$ in the 0.3-1~kHz range (blue trace in Fig.~\ref{fig:dBspectrum}), the presence of the gradient raises the noise level to a value of $\approx$2~pT/$\mathrm{\sqrt{Hz}}$ in the 30-1000~Hz (red trace in Fig.~\ref{fig:dBspectrum}).

In the low frequency range ($<$30~Hz) we observe noise decaying like $\delta{B}\propto 1/f$.
This noise is attributed to ambient field and power supply instabilities and not to the magnetometer performance proper.
Since our scanner is based on a double modulation technique, as described in Ref.~\cite{colombo2016b}, we just need to detect a magnetic signal oscillating in the frequency band around a chosen modulation field frequency.
We have thus proven that we can operate the proposed scanner in the frequency range 30-1000~Hz without loss in sensitivity.

\section{Simulation of the signal and point spread function}\label{ref:signals}
The general idea of our scanner is closely related to the x{--}space variant of MPI \cite{GoodwillAM2012}.
The selection field will saturate all  particles except those located  close to the field free line (FFL).
By adding a harmonically oscillating field $\vec{H}^\mathrm{mod}(t)$ to the selection field $\vec{H}^\mathrm{sel}$, only the unsaturated MNPs will induce a modulation $\delta \vec{B}_\mathrm{mod}$ of the flux density at the sensor position.
The signal of interest is  the amplitude $\delta B_\mathrm{mod}$  of that  field oscillation, which is proportional to the integral contribution of all particles located along the FFL.
When mechanically moving the sample with respect to the FFL, one can thus acquire an image of the MNPs' density distribution.
Conversely to standard MPI techniques our scanner works on the direct detection of the MNP response at the drive frequency.
Detection of higher harmonics can also be envisioned.

The AM, located at $\vec{r}_\mathrm{AM}$, measures variations $\delta{B}_x(t)$ of the field component $B_x$ induced by the magnetization of the MNP sample.
The flux density produced at the sensor location $\vec{r}_\mathrm{AM}$ by a point-like magnetic moment $\vec{\mu}_\mathrm{s}$ located at  $\vec{r}_\mathrm{s}$  is given by 
\begin{equation}
\delta{\vec{B}}=\frac{\mu_0}{4\pi}\left[3\frac{\left[(\vec{r}_\mathrm{AM}-\vec{r}_\mathrm{s})\cdot\vec{\mu}_\mathrm{s}\right](\vec{r}_\mathrm{AM}-\vec{r}_\mathrm{s})}{\left|\vec{r}_\mathrm{AM}-\vec{r}_\mathrm{s}\right|^5}-\frac{\vec{\mu}_\mathrm{s}}{\left|\vec{r}_\mathrm{AM}-\vec{r}_\mathrm{s}\right|^3}\right]\,
\label{eq:dBx}
\end{equation}
and the measured component is given by
\begin{equation}
\delta{B}_x=\delta{\vec{B}}\cdot\hat{x}\,.
\label{eq:componentDec}
\end{equation}
The magnetic moment depends on the local field $\vec{H}(\vec{r}_\mathrm{s})$ and is described -- in  the approximation of a monodisperse MNP suspension -- by the Langevin model function
\begin{equation}
\vec{\mu}_\mathrm{s}(\vec{r}_\mathrm{s})=
\mu_s \hat{H}(\vec{r}_\mathrm{s}) \mathcal{L}\left(\frac{\left|\vec{H}(\vec{r}_\mathrm{s})\right|}{H_k}\right)\,,
\label{eq:muVec}
\end{equation}
with $\mathcal{L}(x){=}\coth(x){-}x^{-1}$ and $\hat{H}(\vec{r}_\mathrm{s}){\equiv}\vec{H}(\vec{r}_\mathrm{s})/\left|\vec{H}(\vec{r}_\mathrm{s})\right|$ denotes the direction of the local field.
The latter is produced by the coil system (selection coils, modulation coils and the corresponding compensation coils) and is thus known.
Since the modulation field is time-dependent it is useful to decompose the local field into selection and modulation components, according to
\begin{equation}
\vec{H}(\vec{r}_\mathrm{s}, t)=\vec{H}^\mathrm{sel}(\vec{r}_\mathrm{s})+\vec{H}^\mathrm{mod}(\vec{r}_\mathrm{s})\cos\left(2\pi f_\mathrm{mod}t\right)\,.
\label{eq:blocDec}
\end{equation}
By inserting \eqref{eq:blocDec} into \eqref{eq:muVec}, and by expanding it in a Fourier series, we get the modulated magnetic moment 
\begin{align}
\vec{\mu}_\mathrm{s}(\vec{r}_\mathrm{s}, t)=\mu_\mathrm{s}\hat{H}(\vec{r}_\mathrm{s},t)\sum_{n\geq 0} m_n\left(\vec{r}_\mathrm{s}\right)\cos(2\pi n f_\mathrm{mod} t)\,,
\end{align}
where the coefficients $m_n$ are given by
\begin{align}
m_0\left(\vec{r}_\mathrm{s}\right)=&\mathcal{L}\left(\frac{\left|\vec{H}^\mathrm{sel}(\vec{r}_\mathrm{s})\right|}{H_k}\right)\\
m_n\left(\vec{r}_\mathrm{s}\right)=&\frac{2}{\pi}\int_0^{\pi}\mathcal{L}\left(\frac{\left|\vec{H}(\vec{r}_\mathrm{s})\right|}{H_k}\right)\cos(2\pi n f_\mathrm{mod} t)\mathrm{d}(2\pi f_\mathrm{mod}t)\nonumber\\
=&\frac{2\left|\vec{H}^\mathrm{mod}(\vec{r}_\mathrm{s})\right|}{\pi H_k}\int_{-1}^{1}\mathcal{L}'\left(\frac{\left|\vec{H}^\mathrm{sel}(\vec{r}_\mathrm{s})+y~\vec{H}^\mathrm{mod}(\vec{r}_\mathrm{s})\right|}{H_k}\right)\nonumber\\
&\phantom{\frac{2\left|\vec{H}^\mathrm{mod}(\vec{r}_\mathrm{s})\right|}{\pi H_k}\int_{-1}^{1}}\sqrt{1-y^2}U_{n-1}(y)~\mathrm{d}y\,,
\end{align}
where $\mathcal{L}'(x)\equiv\mathrm{d}\mathcal{L}(x)/\mathrm{d}x$ and $U_k(x)$ is the $k$-th order Chebyshev polynomial of the second kind.
We note that for $\left|\vec{H}^\mathrm{mod}(\vec{r}_\mathrm{s})\right|{<}H_k$ we have
\begin{align}
m_1\left(\vec{r}_\mathrm{s}\right)&\approx \frac{\left|\vec{H}^\mathrm{mod}(\vec{r}_\mathrm{s})\right|}{H_k}\mathcal{L}'\left(\frac{\left|\vec{H}^\mathrm{sel}(\vec{r}_\mathrm{s})\right|}{H_k}\right)\\
m_{n>1}\left(\vec{r}_\mathrm{s}\right) &\approx ~0
\end{align}
yielding,
\begin{align}
\vec{\mu}_\mathrm{s}(\vec{r}_\mathrm{s}, t)\approx &\mu_\mathrm{s}\hat{H}(\vec{r}_\mathrm{s},t)
\left[
\mathcal{L}\left(\frac{\left|\vec{H}^\mathrm{sel}(\vec{r}_\mathrm{s})\right|}{H_k}\right)\right.+\nonumber\\
&\left.
\frac{\left|\vec{H}^\mathrm{mod}(\vec{r}_\mathrm{s})\right|}{H_k}\mathcal{L}'\left(\frac{\left|\vec{H}^\mathrm{sel}(\vec{r}_\mathrm{s})\right|}{H_k}\right)
\cos(2\pi f_\mathrm{mod} t)
\right]\,.
\label{eq:fulmom}
\end{align}
By inserting Eq.~\eqref{eq:fulmom} into Eq.~\eqref{eq:dBx}, and using Eq.~\eqref{eq:componentDec} yields the detected signal $\delta{B}_x(\vec{r}_\mathrm{s}, t)$.
The system's point spread function (PSF) -- an example of which is shown in Fig.\ref{fig:2dsketch} -- is then obtained by demodulating the latter at the modulation frequency $f_\mathrm{mod}$. 

We point out that the signal contains also higher harmonics of $f_\mathrm{mod}$ since $m_{n>1}{\neq}0$ when $\left|\vec{H}^\mathrm{mod}\right|{\geq}H_k$ which could lead to a better suppression of the  modulation coils' fringe field.
\begin{figure}[!t]
	\centering
	\includegraphics[width=\columnwidth]{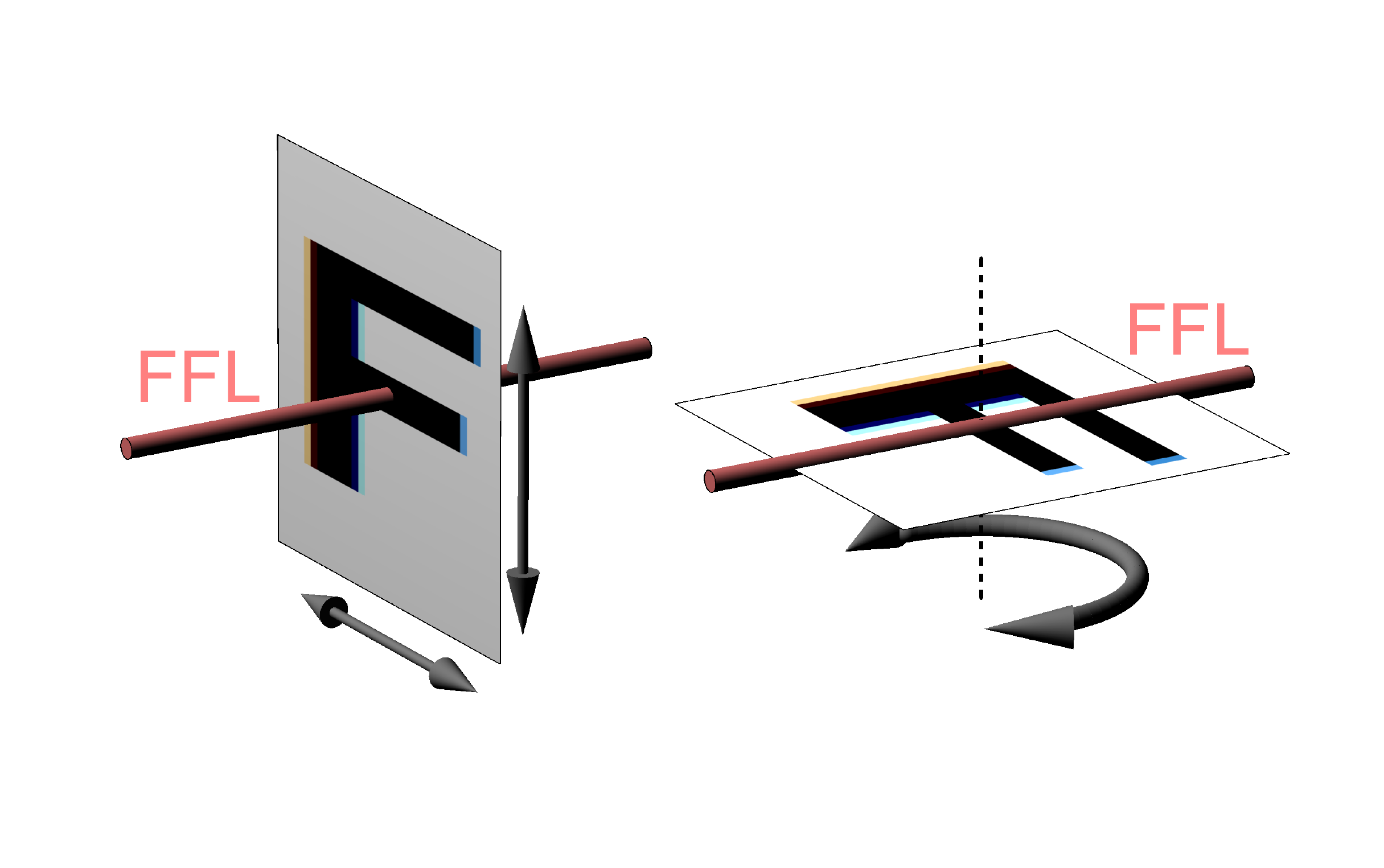}
	\caption{FFP-based (left) and FFL-based (right) 2D scanner sketches. The \textbf{F} symbol represents the MNP distribution in the plane. Pink cylinder represent the FFL. double arrows denote directions of the mechanical motion of the sample.}
	\label{fig:Fplot}
\end{figure}
%
\section{Conclusion}
In this paper we have proposed a novel design for mechanical MPI scanner operating at low frequency ($\leq$~kHz) based on atomic magnetometry.
We have developed a selection coil system which allows to expose the MNP sample to T/m gradient field free line.
The atomic magnetometer measures the flux density $\delta{B}_x{\propto}M$ produced by the MNP's magnetization $M$.
Modulation coils are deployed to extract the magnetic susceptibility d$M(H)/$d$H$ proportional to the MNP density on the FFL.
Corresponding compensation coils reduce, at the magnetometer location, the fringe field and gradients from the selection and modulation coils to sufficiently low values that do not compromise the AM's sensitivity.
In the near future we plan to realize a mechanical 2D scanner, two possible variants of which are illustrated in Fig.\ref{fig:Fplot}.
The mechanical motion implies a rather slow scan-speed.
However, the much lower frequencies than in conventional MPI scanners will make a much broader variety of particles compatible (in particular larger particles) with the MPI method.
%
%
\paragraph{Acknowledgements}
We acknowledge financial support by grant No. $200020~162988$ of the Swiss National Science Foundation


%
\end{document}